# On the performance of some new Multiuser FSO-MIMO Communication Systems

M. A. Amirabadi

Email: m_amirabadi@elec.iust.ac.ir

**Abstract-** The practical implementation of maximum likelihood detection is limited by its high complexity as well as requiring perfect channel state information. Although conventional blind detection techniques reduce complexity, they degrade performance and require blind channel state information. In this paper (for the first time), a deep learning based blind detection and a joint blind detection-constellation shaping structure are presented (to solve this problem). This paper (deeply) goes through the problem and discusses several (practical) scenarios, including single user, multiuser with resource (channel) allocation, and multiuser without resource allocation (multiuser interference). In order to show the universality of the proposed systems, wide atmospheric turbulence regimes, from weak to strong are considered, and single input single output, as well as multi-input multi-output structures are considered. Results indicate that without channel estimation, a deep learning based (blind) detector (despite its very few complexity, and despite conventional systems require it), could have a very favorable performance at all around. So, it is expected that practical implementations of the proposed structures greatly reduce cost, and processing latency, while maintaining performance close enough to the outstanding conventional systems.

**Keywords-** Blind detection, multiuser interference, resource allocation, constellation shaping, free space optical communication, multi-input multi-output, atmospheric turbulence;

## I. Introduction

Free space optical (FSO) communication is one of the most promising approaches for addressing the emerging broadband access market, it has high bandwidth, data rate, and capacity. Its installation is easy and low cost; therefore, it is suitable for high speed applications, disaster recovery, backhaul/ bottleneck of cellular networks. Despite these advantages, FSO system has some barriers, e.g., even in clear weather, the variation of the refractive index of atmosphere turbulence affects its performance, and results in random fluctuation of both intensity and phase of the propagating FSO signal [1]. Usually, FSO setups are implemented in short-range low height links in clear weather, to combat the FSO barriers. Under these circumstances, the only remaining barrier is atmospheric turbulence.

Atmospheric turbulence causes random fluctuations in the received signal intensity [2]. Various statistical distributions have been used to model the effects of atmospheric turbulence, e.g. Lognormal [3], Gamma-Gamma [4], and Negative Exponential [5]. Among these models Gamma-Gamma is highly accompanied by experimental results obtained for weak to strong atmospheric turbulence regimes. The effect of atmospheric turbulence gets worse while considering a multiuser communication link; because the effect of multiuser interference adds up with atmospheric turbulence and leads to higher link degradation. Considering the first problem, use of the multi-input multi-output (MIMO) transceiver structure combined with conventional outstanding detectors such as maximum likelihood with perfect channel estimation could be useful significantly. MIMO uses different received copies of the signal and recovers the original signal better. However, perfect channel estimation has very high computational complexity, and is not preferable for practical applications. Therefore, maximum likelihood with blind channel estimation is more preferable; however, it degrades performance and still requires (blind) channel estimation. Considering the second problem, even the mentioned outstanding solutions could not be appropriate, because the complexity of the problem model increases heuristically, and very large amount of computations are required to have the link remained under connection. Actually, the neighboring user interference is the reason of this problem. However, an available solution for this problem is the resource allocation, the required resource in this problem is the information channels between transmitters, and the receiver.

What would happen if there would be no time/power/space for implementing a huge computational source or solving a heuristic resource allocation? Is there any way for having a favorable performance with low complexity? As a new efficient technique (especially for complex nonlinear models) deep learning has recently taken many considerations towards itself in communication (and especially in optical communications). The aim of deep learning is to enable a system to learn how to do a very complex task from a set of experienced data. Deep learning (in special deep neural network (DNN)) is proven to be very useful, especially as a detector in nonlinear channel models [6]. In addition, the strength of deep learning would be more visual while end to end deploying it in a communication system, because this way it jointly adjust the parameters at the transceiver sides, and uses the potentialities of both sides for improving the performance.

Recently, many investigations considered DNN for fiber OC applications such combating the Fiber effects [7], Modulation Format Identification [8], Optical Performance Monitoring [9], Optical Amplifier Control [10], as well as some Optical Network applications [11]. However, there is no investigation over DNN for wireless optical communication. The reason is that the DNN is better to be implemented anywhere that has a nonlinear model, that the traditional techniques should use more complexity to deal with. Considering machine learning applications in FSO, few basic investigations are developed in applications such as detection [12], distortion correction for sensor-less adaptive optics [13], and demodulator for a turbo-coded orbital angular momentum [14]. All of these works considered machine at the receiver side of a simple FSO-SISO system for a limited (specific) scenario; there is lack of comprehensive investigation in machine (deep) learning for FSO communication.

This paper investigates the effects of multiuser interference and atmospheric turbulence, and presents a DNN based blind detection and a DNN based joint blind detection-constellation shaping structure (as the low complexity efficient solutions for these problems). This paper goes (deeply) through the problem and considers several practical scenarios including single user, multiuser with resource allocation, and multiuser without resource allocation (multiuser interference). In order to have a comprehensive investigation wide atmospheric turbulence regimes, from weak to strong, as well as SISO and MIMO structures are considered. To the best of the authors' knowledge, novelties and contributions of this paper (which are done for the first time in ML for FSO investigations) include:

1) Deploying DNN based blind detection, as well as DNN based joint blind detection-constellation shaping.
2) Considering SISO and MIMO (accordingly equal gain combining (EGC), and selection combining (SC)).
3) Considering single user, multiuser with resource allocation, and multiuser without resource allocation (multiuser interference).
4) This is the most comprehensive work among ML for OC investigations, because it considers wide range of atmospheric turbulence, SISO/ MIMO (SC/ EEG) structures, single user/ multiuser (resource allocation/ multiuser interference) schemes, blind/perfect channel estimation detection, DNN based transmitter/ receiver/ transceiver systems.

The rest of this paper is organized as follows; sections II describes system model. Section III describes the proposed DNN based structures. Section IV is the results and discussions. Sections V is the conclusion of this paper.

**II.    Conventional system model**

The considered conventional multiuser FSO-MIMO system is depicted in Fig.1.a. This system is composed of $N_u$ users, each with $N_t$ transmitter apertures, which want to connect to a base station with $N_r$ receiver apertures. Considering $x_l$, as the generated MQAM symbol at the $l-th; l = 1, ..., N_u$ user, it is passed by the FSO-MIMO channel, and received by an FSO receiver, with $N_r$ receive apertures, in the following form:

$$y_i = R \sum_{j=1}^{N_t} I_{i,j} x_l + n_i, \qquad (1)$$

where $R$ is the photodetector responsivity [15], $n_i$ is the photo-current noise at the $i-th$ receive aperture input noise, including thermal noise, dark current, as well as the shot noise arising from the received signal and the background radiations. While dark current noise can practically be neglected, an additive white Gaussian noise with zero mean and $\sigma^2$ variance can accurately model the sum of these noise sources. $I_{i,j}$ is the atmospheric turbulence intensity of the link between the $j-th$ FSO transmitter of $l-th$ user and $i-th$ receive aperture, which is modeled by Gamma-Gamma distribution by the following probability density function:

$$f(I) = \frac{(\alpha\beta)^{\frac{\alpha+\beta}{2}}}{\Gamma(\alpha)\Gamma(\beta)} I^{\frac{\alpha+\beta}{2}-1} G_{0,2}^{2,0}\left(\alpha\beta I \bigg| \begin{matrix} \cdot, \cdot \\ \frac{\alpha-\beta}{2}, \frac{\beta-\alpha}{2} \end{matrix}\right), \qquad (2)$$

where $G_{p,q}^{m,n}(.|\begin{matrix}\cdots\\\cdots\end{matrix})$ is the Meijer-G function [16, Eq. 07.34.02.0001.01], $\Gamma(.)$ is the Gamma function, $\alpha$ and $\beta$ are the effective number of large-scale and small-scale eddies of scattering environment, respectively, and they are related according to Rytov variance [17].

The conventional structure is not the aim of this paper, and is just used as an outstanding structure to compare the proposed DNN based structures with. The proposed DNN based structures are completely blind, in the sense that they even do not estimate the channel blindly. However, both perfect channel estimation and low complexity blind channel estimation [18] are considered in the conventional system (because, despite DNN based detector, maximum likelihood detector requires either perfect or blind channel estimation). Accordingly, at the receiver (of both conventional and DNN based structures), EGC, and SC combiners are implemented, because the FSO channel is a real valued random variable, and EGC and SC do not require channel information in this situation. After the combiner, at each time slot, there are $N_u$ symbols available at the receiver, but only one of them could be processed by base station.

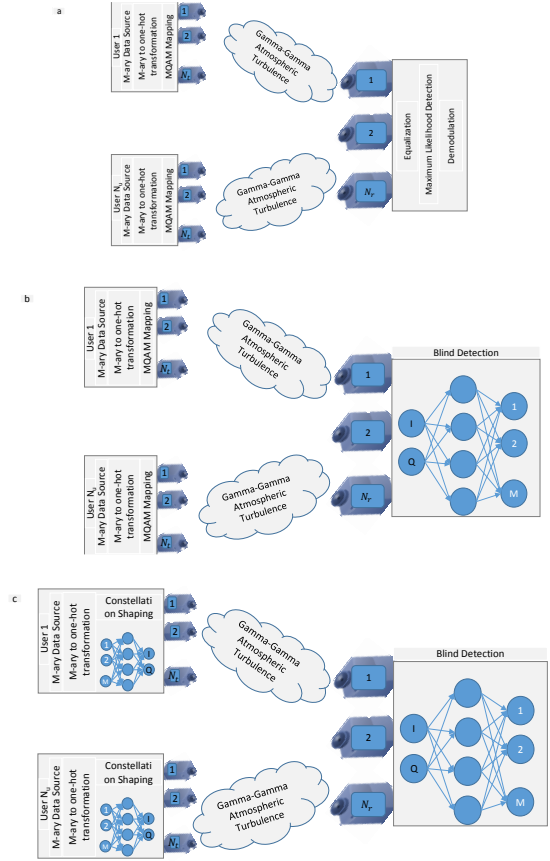

Fig.1. a. Conventional Multi-user FSO-MIMO communication system, b, and c. Proposed DNN based Multi-user FSO-MIMO structures.

There are two assumptions for this situation in this paper including with resource allocation, and without resource allocation (multiuser interference). Considering resource allocation, one of the simplest and best channel allocations is allocating to the user with best channel condition; therefore the receiver selects the signal with the maximum SNR. Considering multiuser interference case, one of the received $N_r$ signals is the target signal and the others are interference signals. Assume $u$ is the index of the target signal. In SC only one of the received apertures is selected (selects the received signal with maximum SNR); assume $p$ is the index of the selected receive aperture by SC, the ML receiver for the SC becomes as follows:

$$\hat{x}_u = \min_{\tilde{x}_u} \left| y - \eta \sum_{j=1}^{N_t} I_{p,j} \tilde{x}_u \right|^2, \tag{3}$$

where $\tilde{x}_u$ is a symbol of the transmitted constellation map. In EGC the received signals are multiplied by the phase conjugate of the channel intensity (to remove channel phase), and then summed. However, FSO channel only takes real values; therefore, in FSO, EGC is only the summation of the received signals, and can be written as follows [19]:

$$y = \sum_{i=1}^{M} y_i = \frac{\eta}{N_t} \sum_{i=1}^{M} \sum_{j=1}^{N_t} I_{i,j} x_u + \sum_{i=1}^{M} n_i. \tag{4}$$

ML Receiver for the EGC will be as follows [19]:

$$\hat{x}_u = \min_{\tilde{x}_u} \left| y - \frac{\eta}{N_t} \sum_{i=1}^{N_r} \sum_{j=1}^{N_t} I_{i,j} \tilde{x}_u \right|^2. \tag{5}$$

### III. Proposed DNN based structures

#### A. Classical transmission/ DNN based blind detection

The proposed Classical transmission/ DNN based blind detection is presented in Fig.1.b. This structure is the same as the conventional structure in section II, except that it has no channel estimation, and uses a DNN for detection. Consider $x_l$ as the generated $M$-ary symbol at the $l-th$ user, this symbol is first converted to a one-hot vector, then mapped on an MQAM constellation (this is because the one-hot vector is required at the output of the DNN, actually it is the label vector required for training the DNN). Then, $N_t$ copies of the mapped symbol are transmitted simultaneously from $N_t$ transmit apertures. The transmitted signal is encountered by Gamma-Gamma atmospheric turbulence, and added by the receiver AWGN with zero mean and $\sigma^2$ variance. Different received copies of the transmitted signals are then combined by either EGC or SC. So at this point, there are $N_u$ complex signals, each related to one of the users, the same as conventional structure, the resource allocation as well as multiuser interference schemes are considered here. The selected signal (or interfered signal in multiuser interference) is then entered a DNN with 2 input neurons (because the input is complex but the DNN accepts real, so real component enters one neuron and imaginary component enters another one), $M$ output neurons, $N_{hid}$ hidden layers, $N_{neu}$ per hidden layer neurons, activation function $\alpha(.)$, weight $W$ and bias $b$. The purpose is to adjust DNN parameters (weight and bias) such that the receiver could better recover original M-ary symbol.

In order to solve this problem efficiently, the DNN should be trained. The first step in training a DNN is selecting and tuning its hyperparameters. In this step the structure of the DNN as well as the specific parameter values should be defined. The structural hyperparameters of DNN include the number of layers, number of neurons, layer type, activation function, loss function, optimizer, learning rate, number of epochs, and batch size. After tuning the inputs of each layer are multiplied by corresponding weights, added by biases, summed, and then passed the activation function. Outputs of each layer are the inputs of the next layer, and the procedure continues until reaching the last layer. Considering the original one-hot vector at the transmitter as $s$ and the output of the one-hot vector as $\hat{s}$, the aim is to reduce the difference between them. Therefore, a loss function should be defined and calculated for each individual transmitted symbol and expected over the whole batch size. The proposed loss function could be defined as follows [20]:

$$L(\boldsymbol{\theta}) = \frac{1}{K} \sum_{k=1}^{K} \left[ l^{(k)}(\boldsymbol{s}, \hat{\boldsymbol{s}}) \right] \tag{6}$$

where $\boldsymbol{\theta}$ is the DNN parameter vector (include weight and bias), $K$ is the batch size, $l(.,.)$ is loss function (in this paper, it is cross entropy function). Then it's turn to minimize the loss function iteratively by the following formulation:

$$\boldsymbol{\theta}^{(m+1)} = \boldsymbol{\theta}^{(m)} - \eta \nabla_{\boldsymbol{\theta}} \tilde{L}(\boldsymbol{\theta}^{(m)}) \tag{7}$$

where $\eta$ is learning rate, $m$ is the training step iteration, and $\nabla_{\boldsymbol{\theta}} \tilde{L}(.)$ is the estimate of the gradient. Actually, the error derivation ($\nabla_{\boldsymbol{\theta}} \tilde{L}(\boldsymbol{\theta}^{(j)})$) is fed back to the DNN as an updating guide; the positive step size is known as the learning rate ($\eta$) [21]. Optimization is a tricky subject, which depends on the input data quality and quantity, model size, and the contents of the weight matrices. Stochastic Gradient Descent methods could be used for determining update direction and solving the above problem. Among them, Adam is the most widely used algorithm for this task [22].

#### B. DNN based blind transmission/ DNN based blind detection

The proposed DNN based transmission/ DNN based blind detection is presented in Fig.1.c. This structure is the same as the proposed conventional transmission/ DNN based blind detection, except that it has a DNN at the transmitter; the aim is to jointly (and blindly) shape the constellation and detect the symbol. Consider $x_l$ as the generated M-ary symbol by $l-th$ user; it is first converted to a one-hot vector. Then entered a DNN with $M$ input and 2 output neurons; for simplicity, and without loss of generality, other DNN hyperparameters are exactly the same as descriptions of in section III.B. Complex summation of the output of this DNN results in a complex number, which stands for the location of the transmitted symbol in the signal constellation. Actually, this DNN has the role of a constellation shaping, such that the effect of atmospheric turbulence on the propagating signal be minimized. Then $N_t$ copies of the mapped signal are transmitted simultaneously trough $N_t$ transmit apertures, encountered by Gamma-Gamma atmospheric turbulence, added by AWGN with zero mean and $\sigma^2$ variance, and finally combined by the EGC or SC at the receiver. Again, the selected signal with best channel conditions (or the interfered signal) enters a DNN (with exactly the same structure as the DNN described in section III.A.). The aim is to adjust the parameters of the proposed structure simultaneously. The training procedure is exactly the same as descriptions of section III.A.

## IV. Simulation results

In this section, performances of the proposed DNN based structures are compared with the outstanding conventional structures. In the conventional structures, MQAM modulation is used at the transmitter, and maximum likelihood with perfect channel estimation is used at the receiver. In order to have a deeper view, each plot is composed of three images, which consider single user, multiuser with resource allocation, as well as multiuser without resource allocation (multiuser interference). Each of the images considers SISO and MIMO structures. In summary, wide range of atmospheric turbulence regimes (from weak to strong), different number of transceiver apertures, different combining schemes, different modulation orders, as well as different transceiver structures are considered, compared, and discussed in this section.

Simulations for conventional structures are done in MATLAB, and simulations for DNN based structures are done in Python, Tensorflow environment, which is super-helpful for DNN. Gamma-Gamma and Gaussian random variables are available in MATLAB and Tensorflow. Table. 1. shows values of the used parameters and hyperparameters in simulations, after manually tuning based on previous knowledge from literature and trial and test. Considering sample size to batch size ratio equal to 4 has better performance in tuning. The layer types are deep artificial neural networks, which are selected due to the application type, this type can approximate any arbitrary, nonlinear, continuous, multidimensional function. There are many activation functions available in Tensorflow, in hyperparameter tuning, Relu shows better performance with lower complexity [22]. The number of hidden neurons depends highly on the degree of nonlinearity and model dimensionality of the model. Highly nonlinear systems require more neurons, while smoother systems require fewer neurons. After designing the DNN, it is the turn of selecting a loss function, cross entropy is mostly used in DNN for OC applications, and therefore is used in this paper. The last step it choosing the optimizer; among the optimizers available in Tensorflow, Adam is the mostly used one and shows better results in the tuning procedure. Tuning indicates that the iteration number of 1000 could bring the required accuracy.

Table.1. Tuned hyperparameters in this paper.

| Hyperparameter | Value |
|---|---|
| Modulation order | 16/4 |
| Number of layers | 4 |
| Number of hidden neurons | 40 |
| Batch size | $2^{64}$ |
| Sample size/batch size | 4 |
| Number of Iterations | 1000 |
| Activation function | Relu |
| Loss | Softmax cross entropy |
| Optimizer | Adam |
| Learning rate | 0.005 |
| Gamma-Gamma Atmospheric turbulence Intensity | Strong ($\alpha = 4.2, \beta = 1.4$)/ Moderate ($\alpha = 4, \beta = 1.9$)/ Weak ($\alpha = 11.6, \beta = 10.1$) |

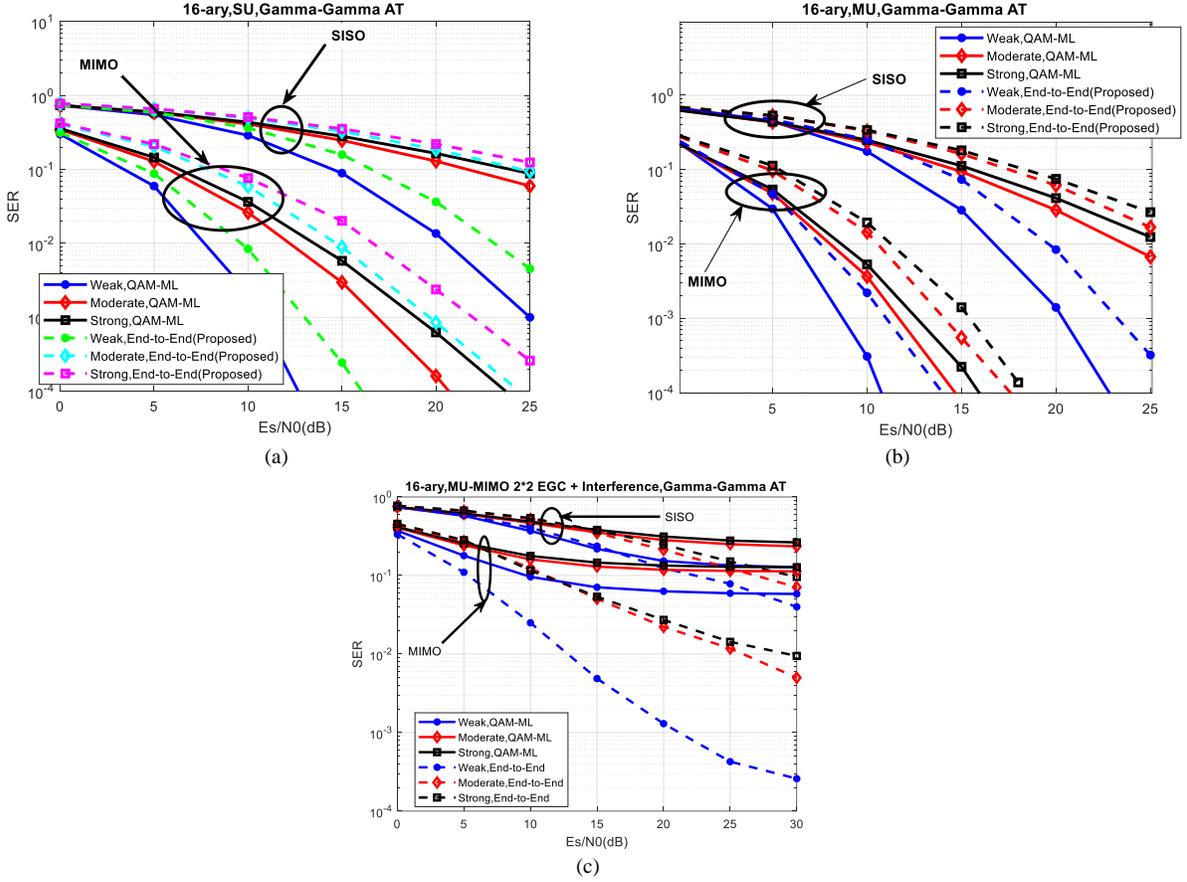

Fig.2. SER of the proposed structure as a function of Es/N0, for different atmospheric turbulence regimes for SISO and MIMO (EGC $2 \times 2$) structures, for a. single user, b. multiuser with resource allocation, and c. multiuser without resource allocation, when modulation order is $M = 16$.

Fig.2. plots SER of the proposed structure as a function of Es/N0, for different atmospheric turbulence regimes for SISO and MIMO (EGC $2 \times 2$) structures, for a. single user, b. multiuser with resource allocation, and c. multiuser without resource allocation, when modulation order is $M = 16$. As can be seen the blind detection performs better, and closer to the state of the art QAM-ML. Accordingly, and considering the fact that the proposed structure performs joint detection and constellation shaping blindly, clarifies the importance of the proposed structure. Because it has reduced complexity, timing and cost greatly (by removing the equalizer part) while almost maintaining the performance. Actually, this is the power of DL that could derive so complicated relationships, and perform a task in the best manner. However, in MIMO scenario this is not such significant, because as mentioned earlier, the most important part in training a DNN is tuning its hyperparameters; in this paper, tuning is done roughly, and once, for the SU-SISO scheme, and all of the other scenarios are developed based on that tuning. However, it could be seen that multiuser scheme performs better than single-user scenario, because in multiuser scheme, the best channel allocation scenario is implemented. In this scenario, the probability that multiple user encounter with bad channel conditions is the multiplication of the same probability for each of them, which decreases significantly while increasing number of users. As can be seen, the difference between blind and perfect CSI cases increase while increasing the Es/N0, because at low Es/N0, noise effect is dominant, and at high Es/N0, channel effects are dominant. In addition this difference is lower for SISO and multiuser cases, because increasing number of transceiver apertures as well as number of users increases number of channels, and in blind situation, increases error probability. Another point that can be seen in the pictures is the difference between blind and perfect CSI in weak, moderate, and strong atmospheric turbulence regime. The reason is that the difference between blind detection and perfect CSI relies on the availability of the channel information itself, not the channel intensity, in other words, the distribution of the channel is the unknown thing in blind detection not the regime.

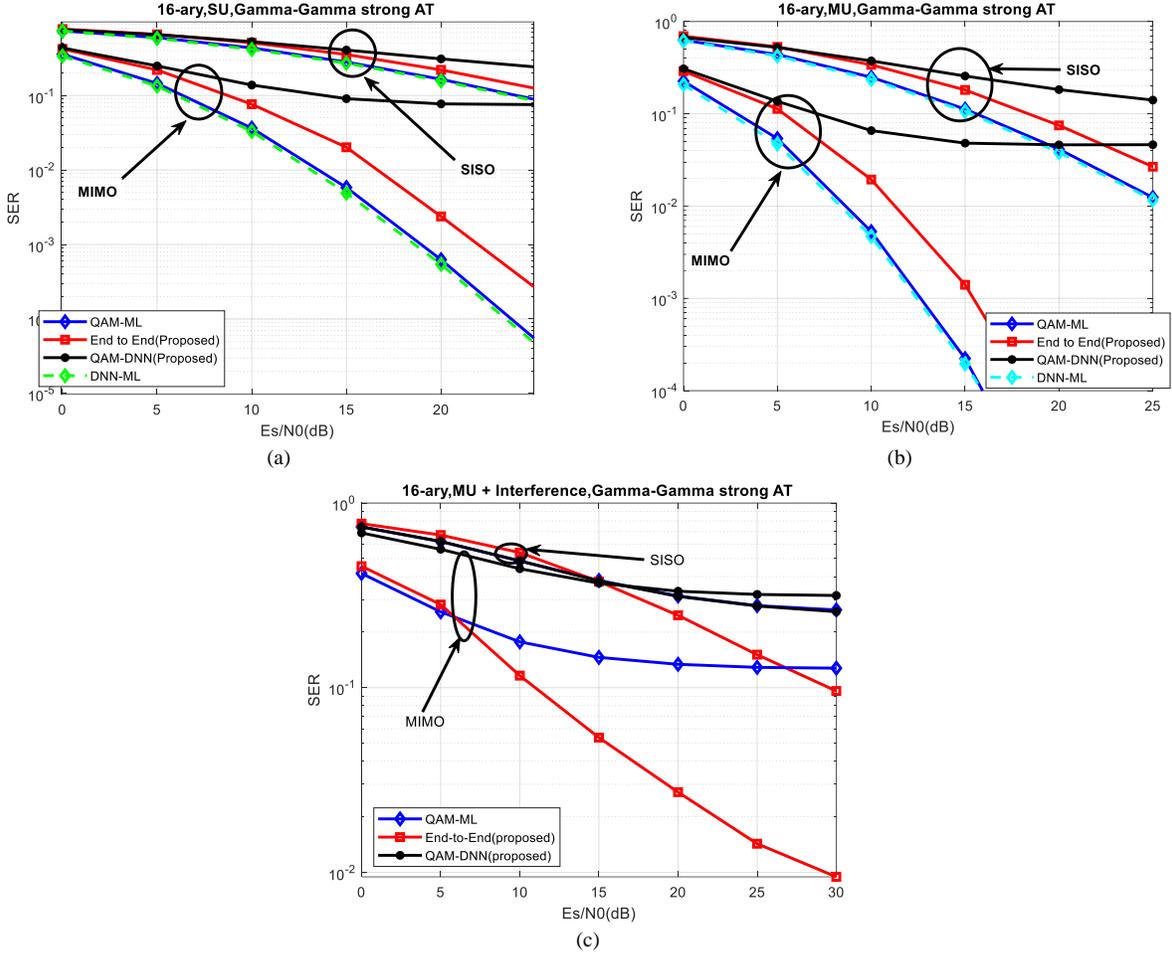

Fig.3. SER of the proposed structures as a function of Es/N0, for SISO and MIMO (EGC $2 \times 2$), for strong atmospheric turbulence, for a. single user, b. multiuser with resource allocation, and c. multiuser without resource allocation, when modulation order is $M = 16$.

In Fig.3, SERs of the proposed structures are plotted as functions of Es/N0, for SISO and MIMO (EGC $2 \times 2$), for strong atmospheric turbulence, for a. single user, b. multiuser with resource allocation, and c. multiuser without resource allocation, when modulation order is $M = 16$. Again, it can be seen that the proposed structures are closer to the state of the art QAM-ML in SU-SISO structure. The effect of constellation shaping clear here; QAM-DNN is the blind detector only, when applying DNN-based constellation shaping (End to End), the performance increases greatly. However, this point should be considered that the proposed ML-based detectors, have somehow more power consumption, because of pilot transmission for channel estimation, and if want to have a fair comparison, the transmission power of DNN-based detectors should be assigned more, because they do not use pilot symbols. As could be seen end to end DNN performs better than end DNN, because it uses the optimum constellation; in other words, in the end to end structure, half of the performance improvement is due to the DNN based detector, and the other half is related to the constellation. The end DNN only has the improvement of implementing the DNN based detector.

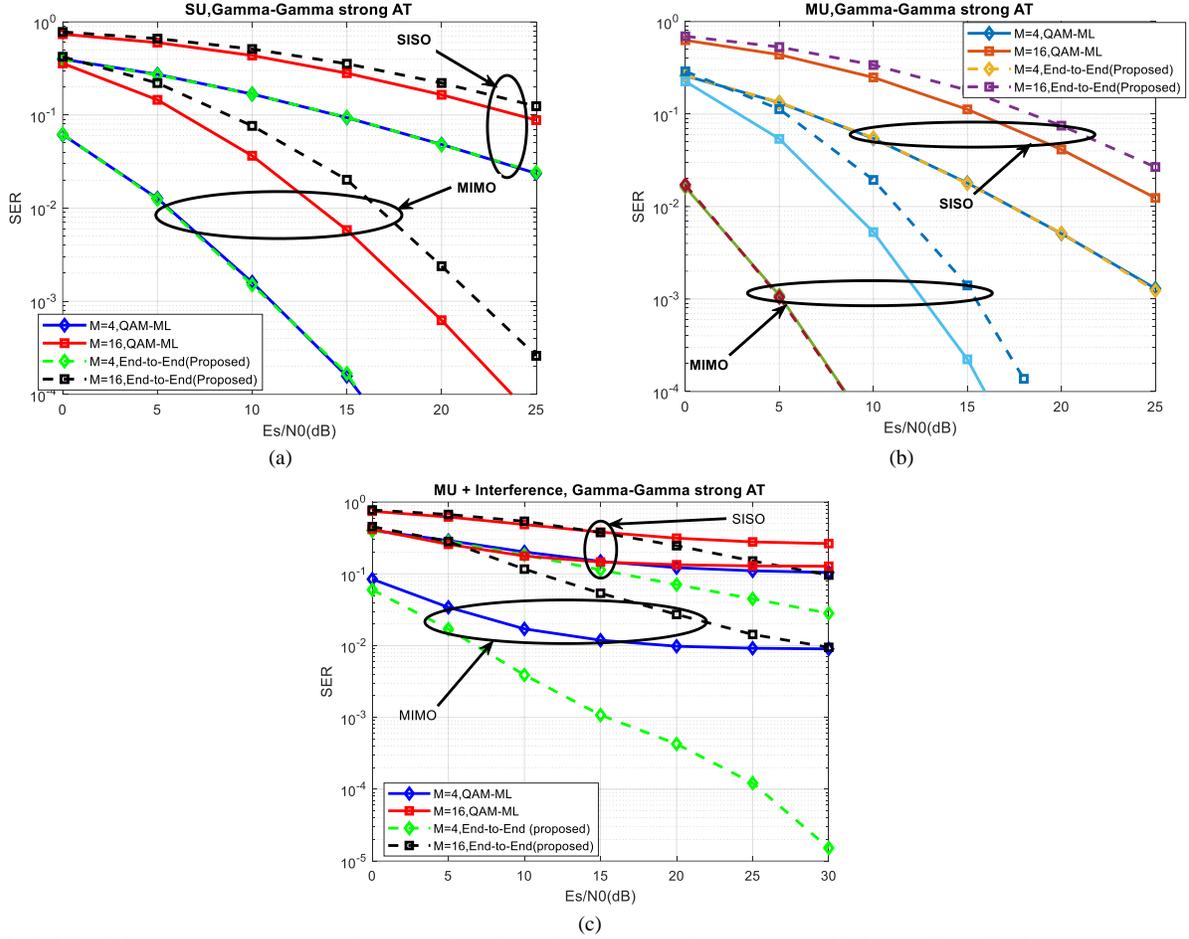

Fig.4. SER of the proposed structure as a function of Es/N0, for different modulation formats, for SISO and MIMO (EGC $2 \times 2$), for strong atmospheric turbulence, for a. single user, b. multiuser with resource allocation, and c. multiuser without resource allocation.

In Fig.4, SER of the proposed structure is plotted as a function of Es/N0, for different modulation formats, for SISO and MIMO (EGC $2 \times 2$), for strong atmospheric turbulence, for a. single user, b. multiuser with resource allocation, and c. multiuser without resource allocation. It can be seen that when modulation order is $M = 4$, the proposed End to End structure achieves the state of the art QAM-ML structure for both SISO and MIMO structures. This is the significant improvement of the proposed structure; it says that either without channel information, end to End DNN could achieve performance of state of the art MQAM-based ML detector. It means that one could reduce power, processing, cost, and latency, while maintaining the performance, in both single/ multiuser and SISO/MIMO structure.

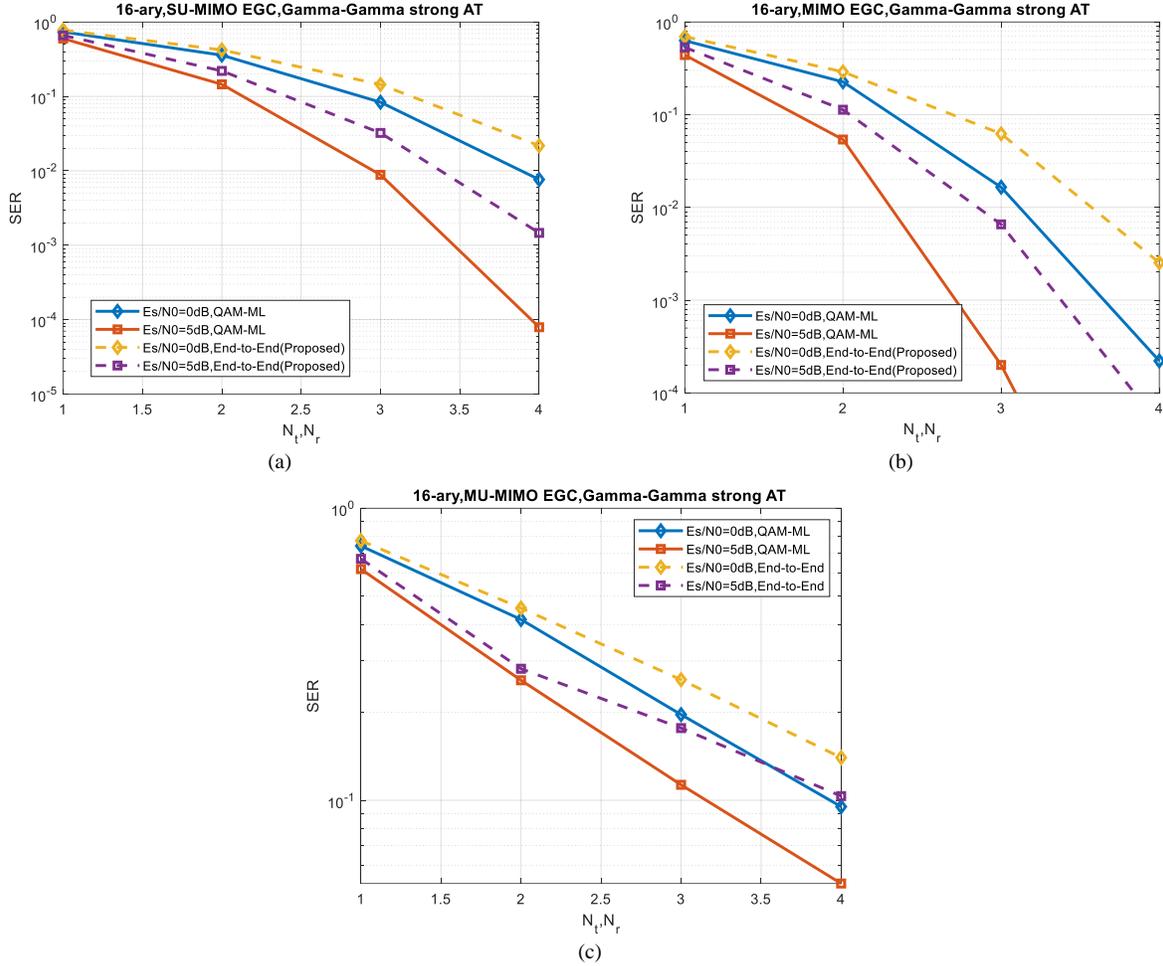

Fig.5. SER of the proposed structure as a function of number of transmitter or receiver antennas, for different Es/N0, for a. single user, b. multiuser with resource allocation, and c. multiuser without resource allocation, when modulation order is $M = 16$.

In Fig.5, SER of the proposed structure is plotted as a function of number of transmitter or receiver antennas, for different Es/N0, for a. single user, b. multiuser with resource allocation, and c. multiuser without resource allocation, when modulation order is $M = 16$.it can be seen that difference between the proposed structure and the state of the art QAM-ML structure increases while increasing the number of transmitters and receivers. The main reason for this maladjustment is that tuning DNN hyperparameters is done based on SISO structure, and the tuned hyperparameters are used for all scenarios. As can be seen, the difference between blind detection and perfect CSI increases while increasing the number of transceiver apertures, because when increasing number of transceiver apertures, number of channel that the blind detector should deal with increases and it is logic to reduce performance in this situation.

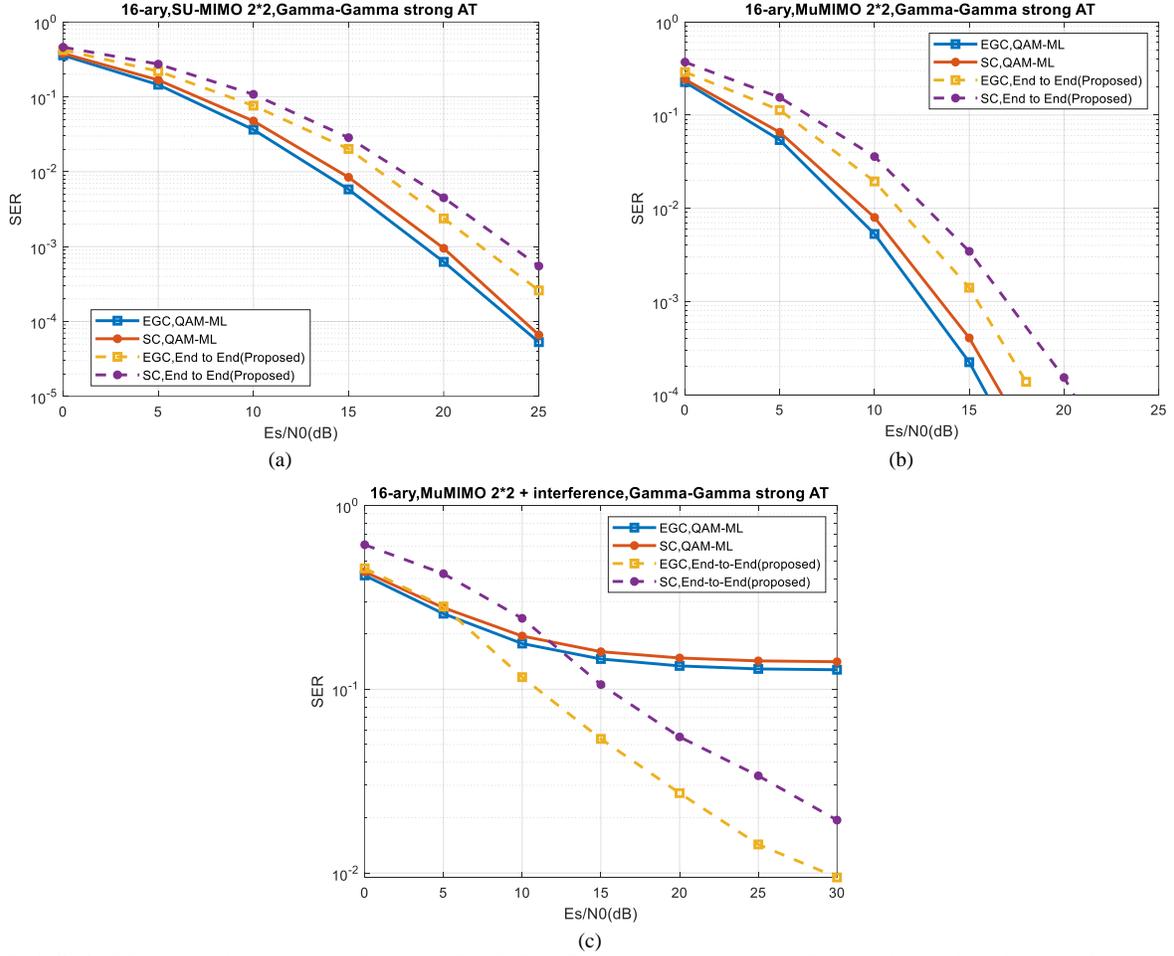

Fig.6. SER of the proposed structure as a function of Es/N0, for different combining schemes, for a. single user, b. multiuser with resource allocation, and c. multiuser without resource allocation, when modulation order is $M = 16$.

In Fig.6, SER of the proposed structure is plotted as a function of Es/N0, for different combining schemes, for a. single user, b. multiuser with resource allocation, and c. multiuser without resource allocation, when modulation order is $M = 16$. Despite the difference mentioned previously, it should be considered that this difference is not so much, and alongside the difference should consider that the proposed structure significantly reduces the processing, complexity, and cost. Another important thing to be considered is the considered scenario, the proposed blind detection scheme got close to the state of the art MQAM-ML in a 16-ary constellation under strong atmospheric turbulence regime. As could be seen, the difference between blind and perfect CSI schemes in both cases of SC, and EGC is the same, the reason is that SC and EGC does not require channel information at all, so existing or non-existing of channel information does not affect them. However, the difference increases in multi-user, because in this scheme increasing number of users, increases number of channels, and because detection is blind, this increases the error probability.

As a pre-final conclusion, in this paper two novel Deep Learning based structures are presented for FSO communication systems. It should be considered that there is lack of investigations in this subject, this work considered simultaneously a wide range of investigations; this is the significance of this work. Considering SISO, MIMO (SC and EGC), single-user, multi-user, weak to strong atmospheric turbulence regimes, all and all in this work show that it is comprehensive. The proposed structure showed that it could be more useful in multiuser scheme, or a few deferring number of apertures would greatly change the performance.

## V.     Conclusion

Maximum likelihood detection has some barriers due to the high complexity as well as requiring perfect channel state information. Considering this problem, this paper (for the first time) presented a deep learning based blind detection and a joint blind detection-constellation shaping structure. This paper (deeply) went through the problem and under wide atmospheric turbulence regime, as well as different single input single output and multi-input multi-output structures, discussed several (practical) scenarios, including single user, multiuser with channel allocation, and multiuser without channel allocation (multiuser interference). The results indicated that despite the fact that the proposed structure has greatly reduced complexity, cost, and latency, it could get enough close to the state of the art ML detection technique.

## References

[1] Amirabadi, M. A., & Vakili, V. T. (2018). Performance comparison of two novel relay-assisted hybrid fso/rf communication systems. IET. COMM.

[2] Amirabadi, M. A., & Vakili, V. T. (2019). A novel hybrid FSO/RF communication system with receive diversity. Optik.

[3] Amirabadi, M. A. (2018). Performance analysis of hybrid FSO/RF communication systems with Alamouti Coding or Antenna Selection. IET JOE.

[4] Amirabadi, M. A., & Vakili, V. T. (2019). On the performance of a multi-user multi-hop hybrid FSO/RF communication system. Optics Communications.

[5] Amirabadi, M. A., & Vakili, V. T. (2018). A new optimization problem in FSO communication system. IEEE Communications Letters, 22(7), 1442-1445.

[6] Jones, R. T., Eriksson, T. A., Yankov, M. P., & Zibar, D. (2018, September). Deep learning of geometric constellation shaping including fiber nonlinearities. In *2018 European Conference on Optical Communication (ECOC)* (pp. 1-3). IEEE.

[7] Giacoumidis, E., Lin, Y., Wei, J., Aldaya, I., Tsokanos, A., & Barry, L. (2019). Harnessing machine learning for fiber-induced nonlinearity mitigation in long-haul coherent optical OFDM. Future Internet, 11(1), 2.

[8] Khan, F. N., Zhong, K., Zhou, X., Al-Arashi, W. H., Yu, C., Lu, C., & Lau, A. P. T. (2017). Joint OSNR monitoring and modulation format identification in digital coherent receivers using deep neural networks. Optics express, 25(15), 17767-17776.

[9] Thrane, J., Wass, J., Piels, M., Diniz, J. C., Jones, R., & Zibar, D. (2017). Machine learning techniques for optical performance monitoring from directly detected PDM-QAM signals. Journal of Lightwave Technology, 35(4), 868-875.

[10] Huang, Y., Gutterman, C. L., Samadi, P., Cho, P. B., Samoud, W., Ware, C., ... & Bergman, K. (2017). Dynamic mitigation of EDFA power excursions with machine learning. Optics express, 25(3), 2245-2258.

[11] Rottondi, C., Barletta, L., Giusti, A., & Tornatore, M. (2018). Machine-learning method for quality of transmission prediction of unestablished lightpaths. IEEE/OSA Journal of Optical Communications and Networking, 10(2), A286-A297.

[12] Zheng, C., Yu, S., & Gu, W. (2015, May). A SVM-based processor for free-space optical communication. In 2015 IEEE 5th International Conference on Electronics Information and Emergency Communication (pp. 30-33). IEEE.

[13] Li, Z., & Zhao, X. (2017). BP artificial neural network based wave front correction for sensor-less free space optics communication. Optics Communications, 385, 219-228.

[14] Tian, Q., Li, Z., Hu, K., Zhu, L., Pan, X., Zhang, Q., ... & Xin, X. (2018). Turbo-coded 16-ary OAM shift keying FSO communication system combining the CNN-based adaptive demodulator. Optics express, 26(21), 27849-27864.

[15] X. Liu, "Free-space optics optimization models for building sway and atmospheric interference using variable wavelength," *IEEE Transactions on Communications*, vol. 57, no. 2, pp. 492–498, 2009.

[16] Wolfram, The Wolfram functions site, Available: http://functions.wolfram.com.

[17] Amirabadi, M. A., & Vakili, V. T. (2019). Performance of a relay-assisted hybrid FSO/RF communication system. *Physical Communication*, 100729.

[18] Amirabadi, M. A., & Vakili, V. T. (2018). Performance evaluation of a novel relay assisted hybrid FSO/RF communication system with receive diversity. *IET OPT*

[19] Bhatnagar, M. R., & Ghassemlooy, Z. (2016). Performance analysis of gamma–gamma fading FSO MIMO links with pointing errors. Journal of Lightwave technology, 34(9), 2158-2169.

[20] Jones, R. T., Eriksson, T. A., Yankov, M. P., & Zibar, D. (2018, September). Deep learning of geometric constellation shaping including fiber nonlinearities. In 2018 European Conference on Optical Communication (ECOC) (pp. 1-3). IEEE.

[21] Wu, X., Jargon, J. A., Skoog, R. A., Paraschis, L., & Willner, A. E. (2009). Applications of artificial neural networks in optical performance monitoring. Journal of Lightwave Technology, 27(16), 3580-3589.

[22] Amirabadi, M. A. (2019). Novel Suboptimal approaches for Hyperparameter Tuning of Deep Neural Network [under the shelf of Optical Communication]. *arXiv preprint arXiv:1907.00036*.